# BEAM-BEAM RESONANCES FOR DIFFERENT COLLISION SCHEMES

D. Shatilov, BINP, Novosibirsk, Russia
for LNF-BINP "Beam-Beam Team" [*]

*Abstract*

One of the main advantages of proposed by P. Raimondi "Crab Waist" collision scheme [1] is a strong suppression of betatron resonances excited by beam-beam interaction. Some qualitative explanations with numerical examples, describing beam-beam resonances for different collision schemes, were given in [2]. This paper can be considered as an "appendix" (additional illustration) to that one. We performed a number of full 2D betatron tune scans (beam-beam simulations) for different collision schemes, so one can easily see how the beam-beam resonances appear and disappear, depending on the colliding conditions.

## INTRODUCTION

Performing a wide range tune scans we cannot avoid a number of serious simplifications. First of all, as our main goal was to investigate beam-beam resonances only, lattice was represented as simple as possible: just a linear 2x2 block-diagonal matrix. We used the same diagonal noise matrix for all working points, providing that the generated emittances (without beam-beam) will be also the same. Of course, this approach is not "realistic", as near the main coupling resonance the vertical emittance must grow. But we simply had no other choice, since all these distortions very much depend on the actual lattice, which we don't know, especially taking into account a huge number of working points tested for each scan (about 40000). On the other hand, there is a clear advantage of such approach: we studied "pure" beam-beam effects without any other nonlinearities, that makes the results clearer and easier to understand.

One more important restriction is bound up with the fact that we performed "weak-strong" simulations. It implies that in the "bad" working points the numbers (luminosity, vertical blowup) are not correct. On the other hand, we don't need exact numbers in the "bad" areas, we need them only in the "good" ones, where blowup is small, and "weak-strong" approach works well there. Besides, we need to know where the "good" and "bad" areas are located in the space of betatron tunes, and "weak-strong" simulations are quite relevant for this purpose.

Also we should mention that Parasitic Crossings and beam tails (lifetime) were not taken into account. Our main concerns were the luminosity and beam core blowup, their dependence on the betatron tunes. The main goal was to illustrate how the resonances excited by beam-beam interaction depend on the colliding conditions (hour-glass, crossing angle, Piwinski angle, Crab Waist). So, the most informative are comparisons of different pictures (scans) and the numbers of maximum luminosities. These comparisons can be not exact in terms of numerical values, but we believe that qualitatively they are quite relevant.

## SET OF PARAMETERS

For the basis we took the SuperB set of parameters of 15.11.2006, electrons being the "strong" beam (7 GeV) and positrons – the "weak" one (4 GeV):

Table 1: Nominal set of parameters

| Horizontal beta | $\beta^*_x$ (mm) | 20 |
|---|---|---|
| Vertical beta | $\beta^*_y$ (mm) | 0.30 |
| Horizontal emittance | $\varepsilon_x$ (nm) | 1.6 |
| Vertical emittance | $\varepsilon_y$ (nm) | 0.004 |
| Bunch length | $\sigma_z$ (mm) | 6 |
| Energy spread | $\sigma_E$ | $10^{-3}$ |
| Synchrotron tune (e+) | $\nu_s$ | 0.02 |
| Damping decrements | $\alpha_{x,y}$ | $1.175 \cdot 10^{-4}$ |
| Circumference | C (m) | 2250 |
| Number of bunches | $N_b$ | 1733 |
| Particles per bunch (e–) | $N_s$ | $3.52 \cdot 10^{10}$ |
| Particles per bunch (e+) | $N_w$ | $6.16 \cdot 10^{10}$ |
| Crossing angle (full) | $\theta$ (mrad) | 34 |
| Piwinski angle | $\phi$ | 18 |
| "Nominal" tune shifts | $\xi_x, \xi_y$ | 1.26, 3.09 |
| "Actual" tune shifts | $\xi_x, \xi_y$ | 0.004, 0.171 |
| Luminosity | L | $10^{36}$ |

The definition of Piwinski angle and basic relations for beam-beam tune shifts are given below:

$$\phi = \frac{\sigma_z}{\sigma_x} tg\left(\frac{\theta}{2}\right) \quad \xi_y \propto \frac{N \cdot \beta_y}{\sigma_x \sigma_y \cdot \sqrt{1+\phi^2}} \quad \xi_x \propto \frac{N}{\varepsilon_x \cdot (1+\phi^2)} \quad (1)$$

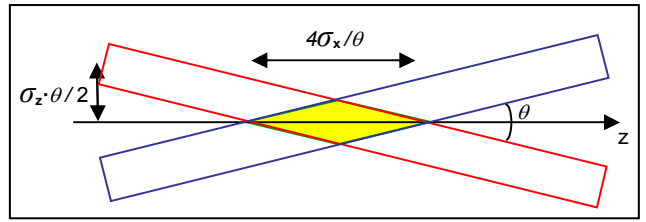

Figure 1: Collision with a crossing angle.

We performed 2D tune scans for the following cases:
- Head-on collisions, different hour-glass: suppressed ($\sigma_z \ll \beta_y$), normal ($\sigma_z = \beta_y$), and enhanced ($\sigma_z \gg \beta_y$).
- Collisions with small Piwinski angle: from 0.2 to 1.2.
- Collisions with large Piwinski angle and different $\beta_y$: large (equal to $\sigma_z$) and small (equal to $\sigma_z/\phi$), with and without Crab Waist.

We tried to keep the nominal set of parameters as close as possible. However, for head-on and small Piwinski angle collisions we had to change some parameters in order to obtain acceptable tune shifts. The idea was to keep the $\xi$ value close to the limit in "good" areas, in this case the pictures of resonances will be the most clear and informative. It should be noted that collisions with changed parameters were not optimized for themselves: we made only minimal changes to get the "correct" $\xi$ value. As we did not take into account PCs, the number of bunches was used only for the total luminosity calculation. We assumed the same $N_b$ in all our simulations, that obviously was very optimistic for head-on and small Piwinski angle collisions. The idea was to compare single bunch luminosities, but renormalized to the total luminosity as for SuperB.

## HEAD-ON AND HOUR-GLASS

First of all, $\beta^*_y$ must be increased by a factor of 20 to match the bunch length. Also, we decided to have the same $\beta^*_x/\beta^*_y$ ratio, the same bunch length and bunch current. If the emittances would be also the same, the "nominal" $\xi_{x,y}$ would not change as well. But we need to reduce them to acceptable values, let's say $\xi_y$=0.07. To achieve this, we increased both emittances by a factor of 44. In this case $\xi_x$=0.0286, and the same crossing angle of 34 mrad would result in Piwinski angle $\phi$ = 0.6. Simulation results for the "nominal" hour-glass ($\sigma_z=\beta_y$) are shown on Fig. 2 (a,b,c). The "geographical map" colors are used there: red corresponds to the maximum luminosity, blue – to the minimum.

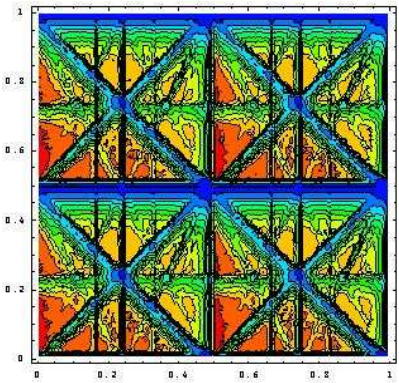

Figure 2a: $\sigma_z=\beta_y$, $L_{max}$ = 2.45·10$^{34}$

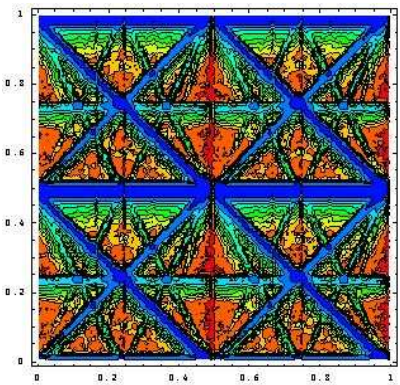

Figure 2b: $\sigma_z=\beta_y$, inverse vertical blowup

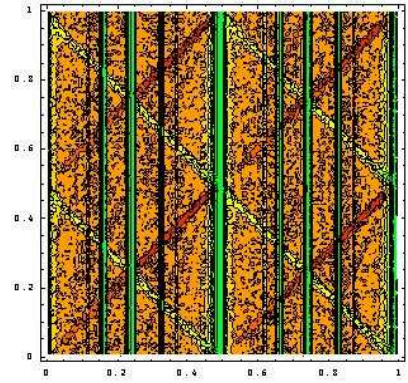

Figure 2c: $\sigma_z=\beta_y$, inverse horizontal blowup

The X-Y betatron resonances appear due to the vertical beam-beam kick's dependence on the horizontal particle's coordinate (amplitude modulation). The horizontal kick also depends on the vertical coordinate, but for the flat beams this dependence is much weaker. The luminosity plot combines both the vertical and horizontal blowups, but for high-order resonances it is better to look at the vertical blowup plot. Resonances L·$\nu_x$ + M·$\nu_y$ = K (L, M – even numbers) are shown on Fig. 3, red lines – up to 4$^{th}$ order, green – 5$^{th}$ and 6$^{th}$ orders. All these resonances are clearly seen on Fig. 2.

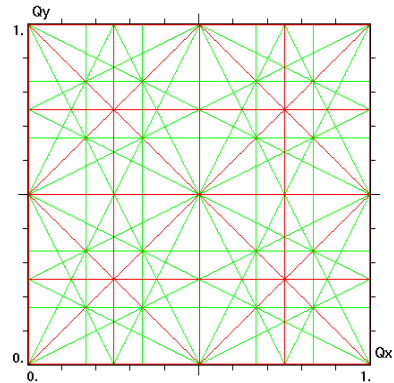

Figure 3: Resonance lines up to 6$^{th}$ order.

Hour-glass effect appears due to Collision Point (CP) longitudinal shift for particles with non-zero Z coordinate. Here CP is a point where a particle meets the center of the opposite bunch, see Fig. 4.

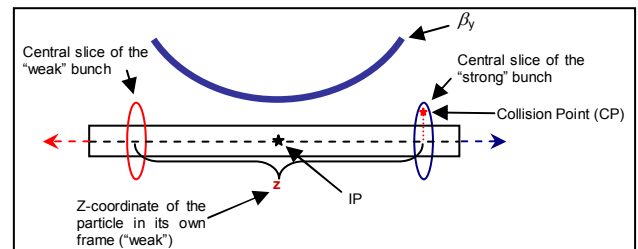

Figure 4: Hour-glass effect

Since $\beta_y$ has a minimum at the IP, $\xi_y$ increases when CP is shifted. Synchro-betatron resonances appear due to the vertical betatron phase modulation at CP and amplitude modulation ($\xi_y$ dependence on CP). Strength of these resonances strongly depends on synchrotron tune: the larger – the worse. On the other hand, the vertical betatron phase averaging over the Interaction Region (IR) results in high-order vertical resonances suppression. Simulation results for suppressed hour-glass are shown on Fig.5 (a,b).

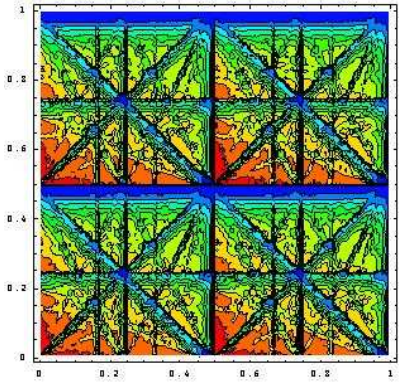

Figure 5a: $\sigma_z = \beta_y/100$, $L_{max} = 3.17 \cdot 10^{34}$

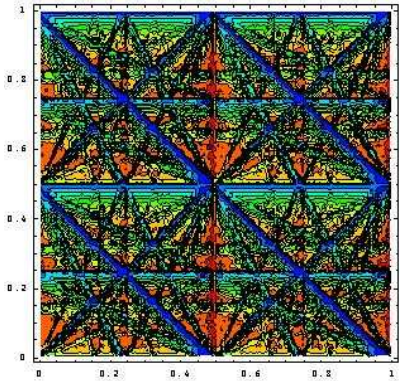

Figure 5b: $\sigma_z = \beta_y/100$, inverse vertical blowup

As we can see, luminosity increases due to geometrical factor and resonance lines become thin, since the synchro-betatron satellites disappeared. On the other hand, more high-order resonances become visible, since the vertical betatron phase averaging disappears, so a particle feels a "solid" kick in a constant phase. Simulations for enhanced hour-glass effect are shown on Fig. 6 (a,b).

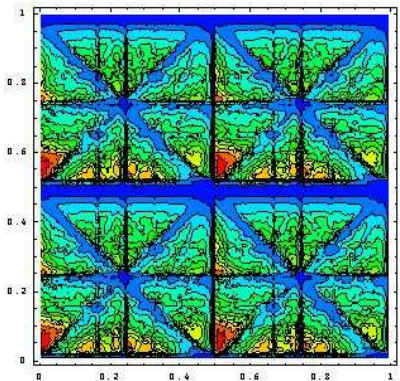

Figure 6a: $\sigma_z = \beta_y \cdot 3$, $L_{max} = 1.62 \cdot 10^{34}$

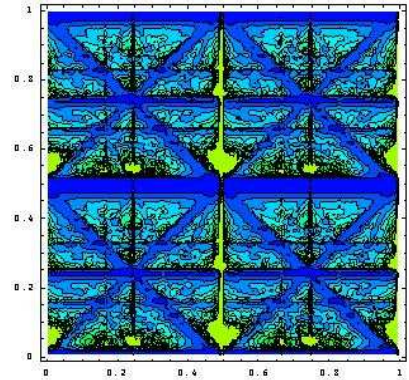

Figure 5b: $\sigma_z = \beta_y \cdot 3$, inverse vertical blowup

Here luminosity decreases due to geometrical factor, and synchro-betatron resonances become much stronger: more satellites, wider resonance lines. So, we cannot find any working point without strong vertical blowup. Taking into account the beam tails, situation looks even worse. Actually it means that the beam-beam tune shift exceeds the limit and must be decreased.

## SMALL PIWINSKI ANGLE

In collisions with a crossing angle the horizontal coordinate of CP (in the strong bunch's coordinate frame) depends on its longitudinal coordinate, see Fig. 6. As a result we obtain amplitude modulation of both horizontal and vertical beam-beam kicks by the particle's synchrotron oscillations, thus exciting strong synchro-betatron resonances. One more important consequence of the crossing angle: it breaks the X-symmetry, so the betatron resonances $L \cdot \nu_x + M \cdot \nu_y = K$ with odd L numbers appear. In particular, low-order resonances $\nu_x \pm 2\nu_y = k$ become very strong.

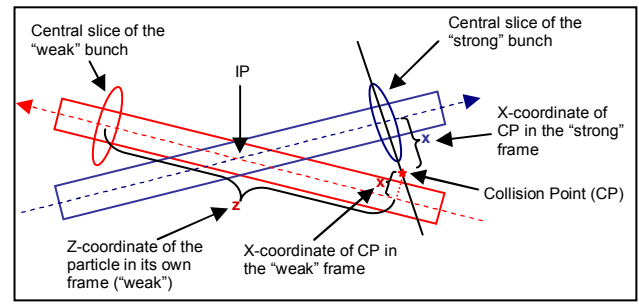

Figure 6: Collision with a crossing angle.

When increasing the crossing angle, luminosity and actual tune shifts decrease due to geometrical factor. Betatron resonances $\nu_x \pm 2\nu_y = k$ become stronger since they need X-asymmetry. On the other hand, "old" betatron resonances (as for head-on) become weaker since the horizontal coordinate of CP (in the strong bunch's coordinate frame) now depends more on the particle's longitudinal coordinate and less on its horizontal betatron coordinate. See the simulation results on Fig. 7 (a, b, c).

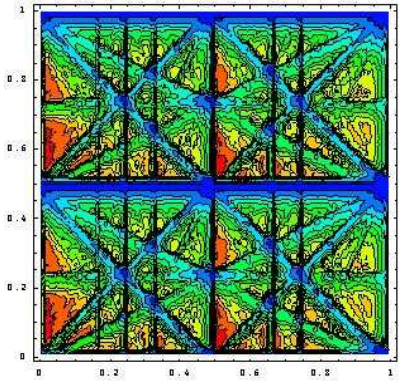

Figure 7a: $\phi = 0.2$, $L_{max} = 2.38 \cdot 10^{34}$

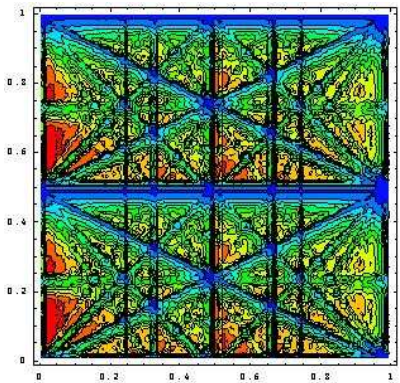

Figure 7b: $\phi = 0.6$, $L_{max} = 2.05 \cdot 10^{34}$

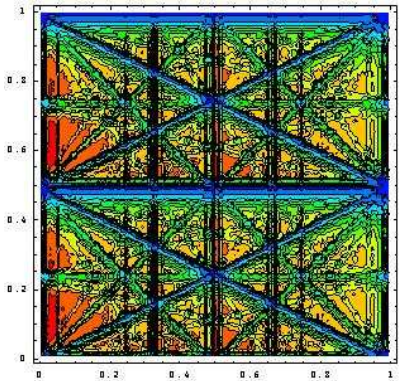

Figure 7c: $\phi = 1.2$, $L_{max} = 1.61 \cdot 10^{34}$

## LARGE PIWINSKI ANGLE

In general, it looks like the larger Piwinski angle – the worse, but for $\phi \gg 1$ we need to change the concept of CP, and this makes a difference. Indeed, for large horizontal separations (in units of $\sigma_x$) the vertical beam-beam kick drops as $1/R^2$, while the horizontal one drops as $1/R$. It means that for the vertical kick the center of the opposite bunch becomes not so important and can be not seen at all by the particles with large longitudinal displacements due to large horizontal separation. Thus CP has to be defined in a different way: it is *the point where a test particle crosses the longitudinal axis of the opposite beam*. In particular it means that the X-coordinate of CP in the "strong" frame is always zero, by the definition.

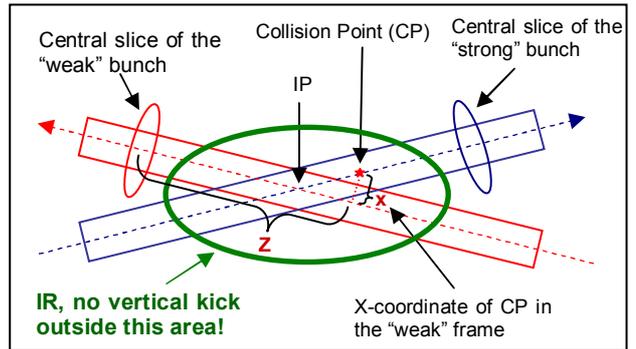

Figure 8: Collision with large Piwinski angle.

Now we simply return back to small emittances and $\beta_x$, as specified in Table 1, thus obtaining Piwinski angle $\phi=18$. For the beginning we did not change $\beta_y$ and keep it equal to the bunch length – just to see the effect of the new CP concept. However, we decreased the bunch current by a factor of ten in order to keep acceptable tune shifts. Since the distance between IP and CP is negligible as compared to $\beta_y$, the vertical beam-beam kick's dependence on the particle's X-coordinate becomes very small. This makes X-Y betatron resonances much weaker than even in the ordinary head-on collisions, see Fig. 9.

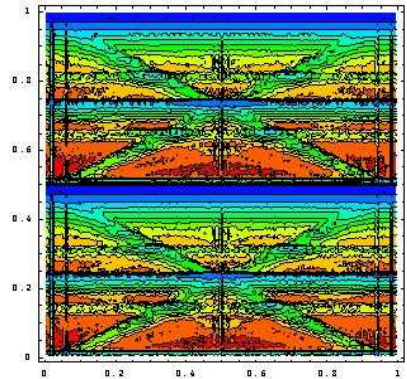

Figure 9: Luminosity, $\phi=18$, $\beta_y = \sigma_z$

The next step is decreasing the $\beta_y$ to its "nominal" value (see Table 1), to fit the overlapping area. Since the shift of CP due to X-betatron oscillations becomes now comparable with $\beta_y$, the vertical betatron phase at CP and $\xi_y$ are strongly modulated, thus exciting X-Y betatron resonances again. Actually, this is rather similar to synchro-betatron resonances exited by the hour-glass effect in head-on collisions. Simulation results are shown on Fig. 10. Here all the parameters are the same as listed in Table 1, except the bunch current which was reduced by a factor of 2.5 to get acceptable tune shifts (without Crab Waist). If we compare Figures 7b and 10, they look rather similar, but there are some differences. In the case of large $\phi$ the horizontal synchro-betatron resonances are enhanced, while the vertical ones are suppressed [4], as well as horizontal betatron ones. As for X-Y betatron resonances, their strength and width also changed, since the sources are different. For small $\phi$ it is an amplitude modulation of the vertical kick, coming from its dependence on X-coordinate. For large $\phi$ the main source is the Y-betatron phase modulation, plus amplitude modulation coming from $\xi_y$ dependence on X-coordinate (due to hour-glass).

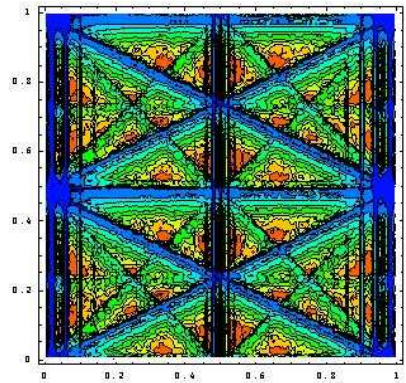

Figure 10: $\beta_y = 0.3$ mm, $L_{max} = 1.6 \cdot 10^{35}$

Finally, we introduce the Crab Waist (CW), which kills the vertical betatron phase modulation. According to [2]

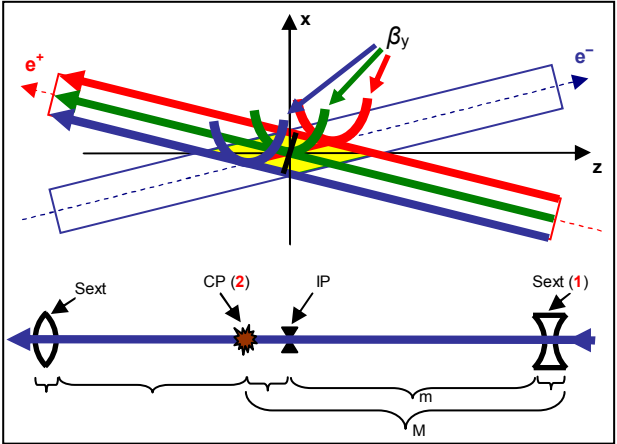

Figure 11: Crab Waist scheme.

the transport matrix $M$ (see Fig. 11) from the entrance of the first sextupole (point 1) to the CP (point 2), vertical betatron motion only, can be written as:

$$M = \begin{pmatrix} 1 & L \\ 0 & 1 \end{pmatrix} \cdot \begin{pmatrix} m_{11} & m_{12} \\ m_{21} & m_{22} \end{pmatrix} \cdot \begin{pmatrix} 1 & 0 \\ V & 1 \end{pmatrix} \quad (2)$$

where the first matrix corresponds to the drift space from IP to CP, L being the drift length, the last matrix corresponds to the sextupole, considered here as a thin linear lens, and in the middle is the unperturbed matrix $m$ from the sextupole location to the IP. For this unperturbed matrix we have $m_{22} = 0$, since $\alpha_y = 0$ at the IP and $\Delta\mu_y = \pi/2$. As a result we get $M_{22} = 0$ as well. On the other hand, considering the "new" lattice (sextupoles included) we can write the standard formula for $M_{22}$:

$$M_{22} = \sqrt{\beta_y/\beta_{1y}} \cdot \left( \cos(\Delta\mu_{1y}) - \alpha_{1y} \cdot \sin(\Delta\mu_{1y}) \right) \quad (3)$$

where $\beta_{1y}$ and $\alpha_{1y}$ are the beta- and alpha-functions at the CP. Since it is the waist at the CP, $\alpha_{1y}$ must be equal to zero, so we get $\cos(\Delta\mu_{1y}) = 0$, resulting in $\Delta\mu_{1y} = \pi/2$, that is exactly what we wanted. In the other words, the vertical betatron phase advance from the first sextupole to CP and then from CP to the second sextupole remains to be $\pi/2$ for all the particles independently on their X-coordinate. This feature allows increasing the beam-beam tune shift by a factor of about 2.5! Thus we return to the nominal bunch current (see Table 1). Simulation results for the nominal waist rotation are shown on Fig. 12.

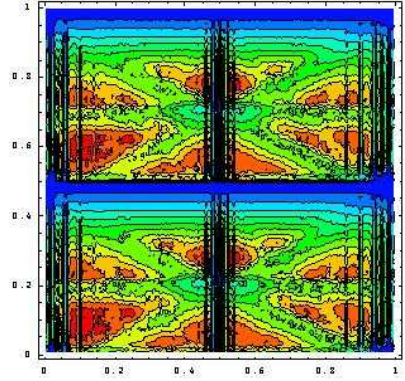

Figure 12: CW = 1, $L_{max} = 1.03 \cdot 10^{36}$

Now let us consider an amplitude modulation of the vertical beam-beam kick caused by the $\beta_y$ modulation at the CP. The vertical tune shift depends on both "weak" and "strong" betas, as follows:

$$\xi_y \propto \frac{\beta_{yw}}{\sqrt{\varepsilon_{ys} \cdot \beta_{ys}}} \quad (4)$$

Here in the numerator we have "weak" $\beta_y$, and in the denominator – "strong" beam size. Without Crab Waist both betas at the CP are actually the same, the difference is negligible when $\theta \ll 1$. It means that $\xi_y$ scales as $(\beta_{ys})^{1/2}$. In the CW scheme $\beta_{yw}$ = const at the CP, so $\xi_y$ scales as $(\beta_{ys})^{-1/2}$, that is inverse dependence of the one without CW, see Fig. 13. This means that if the waist rotation is smaller than the nominal value, the amplitude modulation should decrease while some phase modulation appears again. From here we can conclude that there is some optimum waist rotation angle, as a compromise between amplitude and phase modulations, which should depend on the other parameters ($\xi$, $\phi$, etc.). Usually the optimum lies somewhere in the range of 0.6 to 0.8 of the nominal value.

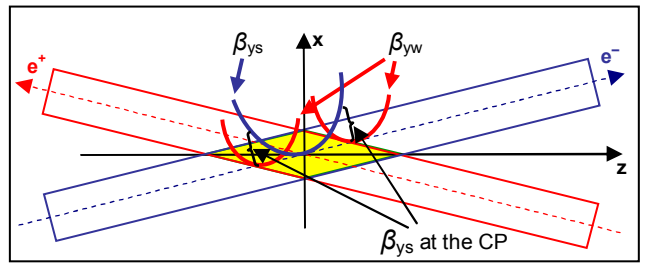

Figure 13: "weak" and "strong" betas at the CP with CW.

In order to check the Crab Waist idea we performed special simulations without $\beta_{ys}$ modulations. This was achieved by increasing $\beta_{ys}$ by a factor of 100 and decreasing the "strong" vertical emittance $\varepsilon_{ys}$ by the same factor, so the vertical beam size was not changed. In these conditions the optimum waist rotation must be shifted to the nominal value, and it was completely confirmed by our simulations. These rather specific simulations, of course, were not realistic, as the "weak" and "strong" beam parameters were very different. The only goal was to demonstrate how the X-Y betatron resonances are suppressed by the Crab Waist. The luminosity tune scans for these conditions without and with Crab Waist are shown on Fig. 14 (a, b).

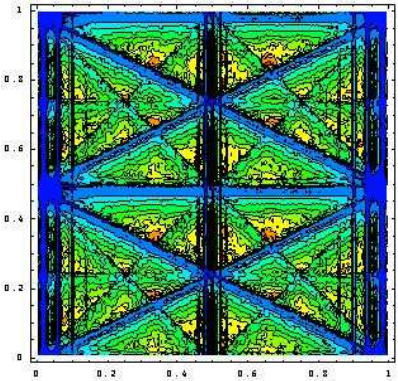

Figure 14a: CW=0, without $\beta_{ys}$ modulations.

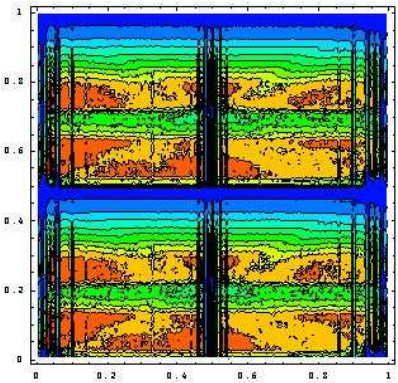

Figure 14b: CW=1, without $\beta_{ys}$ modulations.

As one can see, without Crab Waist removing the $\beta_{ys}$ modulations did not help at all, but with Crab Waist it results in actual vanishing of all X-Y resonances. We still can see the resonances $\nu_x \pm 2\nu_y = k$, but they became rather weak (note the color!). Though they look wide, it is simply due to a very large tune shift: $\xi_y = 0.17$. Also, this is the reason of "shifting" the resonances down.

Finally, we performed a tune scan for the nominal set of parameters with the optimal waist rotation, see Fig. 15. The optimal CW value can be recognized even clearer when performing the beam tails simulation [2]. As for the luminosity scan, the resonances $\nu_x \pm 2\nu_y = k$ become more emphasized for CW=0.8 (Fig. 15) as compared to CW=1 (Fig. 12), but on the other hand the "good" areas become larger for CW=0.8, especially the ones close to half-integer resonances.

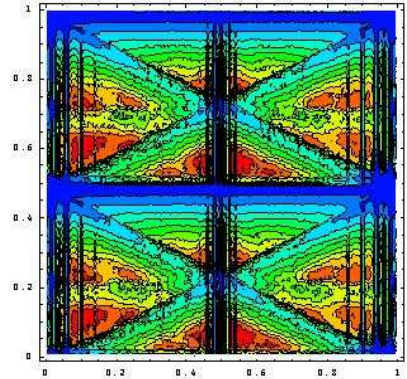

Figure 15: CW=0.8, $L_{max} = 1.05 \cdot 10^{36}$

## CONCLUSIONS

We performed a number of beam-beam simulations for different collision schemes. The main sources of beam-beam resonances which affect the equilibrium particles distribution were recognized, and the luminosity tune scans allowed their clear visualization and identification.

The collision scheme with large Piwinski angle and Crab Waist looks the most promising, since it makes the X-Y modulations much smaller as compared to head-on collision scheme, thus the beam-beam limit $\xi_y$ can be significantly increased, that was confirmed by the recent experimental results obtained on DAFNE [5].

## ACKNOWLEDGEMENTS

All these simulations were performed on **lxcalc** cluster (LNF-INFN, Frascati, Italy). Thanks to LNF Computing Division for their support!